\documentclass[10pt, fleqn]{article}
\usepackage{graphicx}
\usepackage[fleqn]{amsmath}
\usepackage[version=4]{mhchem}
\usepackage{siunitx}
\usepackage{longtable,tabularx}
\usepackage{nccmath}
\usepackage{amsmath}
\usepackage[superscript]{cite}
\usepackage[margin=1in]{geometry}
\usepackage[toc,page]{appendix}
\usepackage{setspace}
\usepackage{amssymb}
\usepackage{tabu}
\usepackage{listings}
\usepackage{afterpage}
\usepackage{titlesec}
\usepackage[labelfont=bf]{caption}
\usepackage{fancyhdr}
\usepackage{pgf}
\usepackage{float}
\usepackage{subcaption}
\usepackage{cleveref}
\usepackage{textcomp}
\usepackage{pict2e}
\usepackage{wasysym}
\usepackage{multicol}
\usepackage{lipsum}
\usepackage{indentfirst}
\newenvironment{Figure}
  {\par\medskip\noindent\minipage{\linewidth}}
  {\endminipage\par\medskip}

\titleformat{\section}
{\normalfont\normalsize\bfseries}{\thesection}{0.5em}{}
\titleformat{\subsection}
{\normalfont\normalsize\bfseries\em}{\thesubsection}{0.5em}{}
\titleformat{\subsubsection}
{\normalfont\normalsize\bfseries\em}{\thesubsubsection}{0.5em}{}
\begin{document}

% Input the sections of text here
\begin{flushright}
\Large % (14 point font)

% Input the paper number: 
\textbf{SSC26-RAIII-03}
\end{flushright}
\begin{centering}      
\large % (12 point font)

% Input the title: 
\textbf{The Solar Neutrino and Astro-Particle PhYsics (SNAPPY) CubeSat Development}\\
\vspace{0.5cm}
\normalsize % set to 10 point font

% Input the author information:
Brian M. Sutin\\
{Skewray Research, LLC}\\
{433 Harrison Ave, Claremont, CA 91711 USA}\\
{smallsat@skewray.com}\\ 
\vspace{0.5cm}

{Edward Bierens}, Brian Doty, Atri Dutta, Jonathan Folkerts, Brooks Hartsock, Kyle Messick, Holger Meyer, Daniel Reichart, Nick Solomey\\
{Wichita State University}\\
{1845 Fairmount Street, Wichita, KS 67260 USA}\\
\vspace{0.5cm}

{Mark Crystal}, Miguel Rodriguez-Otero, Evgeny Kuznetsov\\
{NASA Marshall Space Flight Center}\\
{Martin Rd SW, Huntsville, AL 35808 USA}\\
\vspace{0.5cm}

Robert McTaggart\\
{South Dakota State University}\\
{1055 Campanile Ave, Brookings, SD 57007 USA}\\
\vspace{0.5cm}

James Cutler\\
{University of Michigan}\\
{1320 Beal Avenue, Ann Arbor, MI 48109 USA}\\
\vspace{0.5cm}

{Joel Steinkraus}, Jose G. Rivera\\
{Caltech/Jet Propulsion Laboratory}\\
{4800 Oak Grove Dr, Pasadena, CA 91109}\\

\end{centering}

% Input the abstract: 
\begin{centering}
    \vspace{0.5cm}
    \centerline{\textbf{ABSTRACT}}
    \vspace{0.3cm}
\end{centering}

The SNAPPY CubeSat, which was launched May 3, 2026, will demonstrate and space qualify the $\nu$Sol neutrino-detection technology. The NASA Innovative Advanced Concepts (NIAC) program funded the $\nu$Sol study to determine the scientific and technical challenges associated with a solar neutrino detector operating in close orbit around the Sun, such as a Parker Solar Probe–like orbit at 7 solar radii and potentially as close as 3 solar radii. This program began in 2017 and has continued with funding from NIAC and NASA’s Heliophysics Division. The $\nu$Sol technology detects solar neutrinos using a gallium isotope which decays by emitting two particles spaced apart in time; this allows differentiating neutrino events from cosmic rays.

In the Phase II project review in 2021, concept and science were determined to be feasible; however, two precursor studies were recommended before pursuing a full mission study. These studies were to characterize the true deep-space background for the detector’s gallium double-pulse signal and to collect a statistically significant number of double-pulse events demonstrating that fast electronics can reliably select and analyze this signal. To test double-pulse signals in space, a NIAC Phase III funded building a 3U CubeSat carrying a 0.1-kg gallium-aluminum-gadolinium-garnet (GAGG) detector housed within an active veto array and shielding. Because the detector requires deep-space-like conditions, the CubeSat is designed for a polar low-Earth orbit at 450 km or higher altitude, collecting data over the Earth’s poles above the Van Allen belts.

The spacecraft is built on a NanoAvionics platform. The detector was developed by the Wichita State University Radiation Detector Laboratory, with custom readout electronics designed and built at MSFC. JPL oversaw the mechanical and thermal aspects of the detector and its electronics. The NanoAvionics procurement included the 3U frame, flight and payload computers, power system with rechargeable batteries and deployable solar panels, UHF and S-band transceivers, Sun sensors, GPS, and an avionics card using magnetorquers, as no camera-based pointing stability is required beyond solar panel orientation in a sun-synchronous orbit. We report on the design and construction of this CubeSat for solar neutrino detector background rate measurement.

Because the detector is highly sensitive, with roughly 7\% energy resolution, active veto shielding, and passive shielding using a patented tungsten-powder and epoxy mixture that disintegrates upon atmospheric reentry, SNAPPY enables additional science during the extended mission phase of year two operations. These include measurements of solar wind particle density and energy spectra with particle identification of electrons, protons, and alpha particles; detection of very low-energy gamma rays from galactic gamma-ray bursts without directionality; and a collaboration with amateur radio citizen scientists studying correlations between radio disruptions and solar wind particle density, energy, and species in Earth’s upper atmosphere. We will report on these science studies as well as the main mission of the solar neutrino background rate measurement.

\begin{multicols*}{2}
%Input body text sections here in order

\section{Science of the $\nu$SOL Project}
Neutrinos are a type of subatomic particle produced in radioactive decays of elements naturally occurring on Earth, in fission reactions in nuclear reactors, and in the decays of the nuclei produced in fusion reactions in the sun and other stars. Particle physicists study neutrinos in large ground-based experiments to better understand their fundamental properties but the $\nu$SOL probe is unique as an opportunity to do this in space close to the sun where the flux is around 1000 times greater. Solar neutrinos can also provide knowledge of the fusion reactions in the sun and have been detected on earth since the 1960s \cite{BahcallDavis1976}. 

Measurements of solar neutrinos in space may be able to provide information inaccessible to ground based detection on earth. Viewed (with neutrinos) from earth, the solar fusion core looks basically like a point source. Being close to the sun enables better angular resolution of the solar fusion core. Also, solar neutrinos undergo a transition from a coherent to a decoherent state close to the sun, long before they reach earth. This transition cannot be observed from earth but could be mapped out from a spacecraft.

The $\nu$SOL (Neutrino Solar Orbiting Laboratory) project aims to eventually create a solar neutrino detector probe and send it close to the sun, at least matching the Parker Solar Probe orbit. One of the stepping stones of this mission involves operating a neutrino detector like the one envisioned for $\nu$SOL in space to characterize the backgrounds to the real neutrino signal. The SNAPPY CubeSat mission is capable of doing that and more. We first discuss the main science objective of SNAPPY: to qualify a neutrino detection technique in space for a solar probe mission in the future. Later sections will examine secondary science objectives of the SNAPPY mission.

\section{Technology choice for a solar neutrino detector in space}

The neutrino detection technique for $\nu$SOL must work in a space environment with its background of cosmic rays. Whereas terrestrial detectors can be placed in mines deep underground to shield the cosmic rays, a spacecraft based detector does not have that option. Instead, the detector must use a signal that is unlikely to occur from random backgrounds such as a delayed double pulse, our team has analyzed the noise reduction efficiency of solar neutrinos \cite{NoiseReduction}. This neutrino detection signal is a delayed coincidence measurement between the energy deposited by an initial prompt electron and the energy deposited by a delayed gamma ray produced from nuclear de-excitation. The Feynman diagram in figure \ref{FIG:Feyn} depicts this interaction. 

\begin{Figure}
    \centering
    \includegraphics[width=0.85\linewidth]{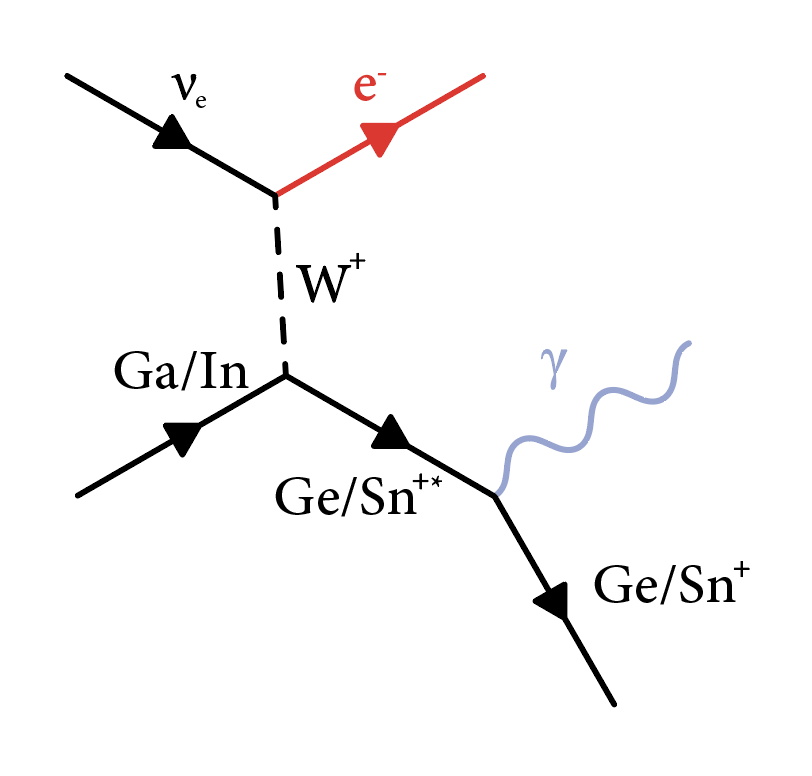}
    \captionof{figure}{Solar neutrino interaction on Ga-71 nucleus. The neutrino interaction produces an electron (always) and an excited state Ge-71 nucleus (with some proportion given by nuclear physics).}
    \label{FIG:Feyn}
\end{Figure}

A solar energy neutrino (with energy from 405 keV to 18 MeV) is absorbed by a Ga-71 nucleus in a Charged-Current (CC) interaction \cite{HAXTON2025139259}. This promotes Ga-71 into Ge-71 in an excited state with a known branching ratio for the given incoming neutrino energy. This excite state decays via a series of gamma ray emissions. Most of these decays have a half-life of less than 1 ns, which is faster than the electronics and the crystal decay time. However, the first excited state of Ge-71 has a half-life of 81 ns \cite{ABUSALEEM2011133}. This combination of a prompt electron signal and a delayed gamma signal, a double-pulse, is distinct from single-pulse signals caused by cosmic rays traversing the detector.

The gadolinium aluminum gallium garnet (GAGG), shown in figure \ref{FIG:GAGG}, is a scintillator that is 22.6\% gallium by mass. Naturally abundant gallium is 39.9\% Ga-71 by mole fraction (39.2\% by mass ), making the whole GAGG crystal 8.86\% Ga-71 by mass, the highest of any known solid scintillator besides the nascent gallium (III) oxide. The electrons and delayed gammas from neutrino interactions cause pulses of light in the crystal that can be detected with photo sensors and digitized with custom electronics.

\begin{Figure}
    \centering
    \includegraphics[width=0.85\linewidth]{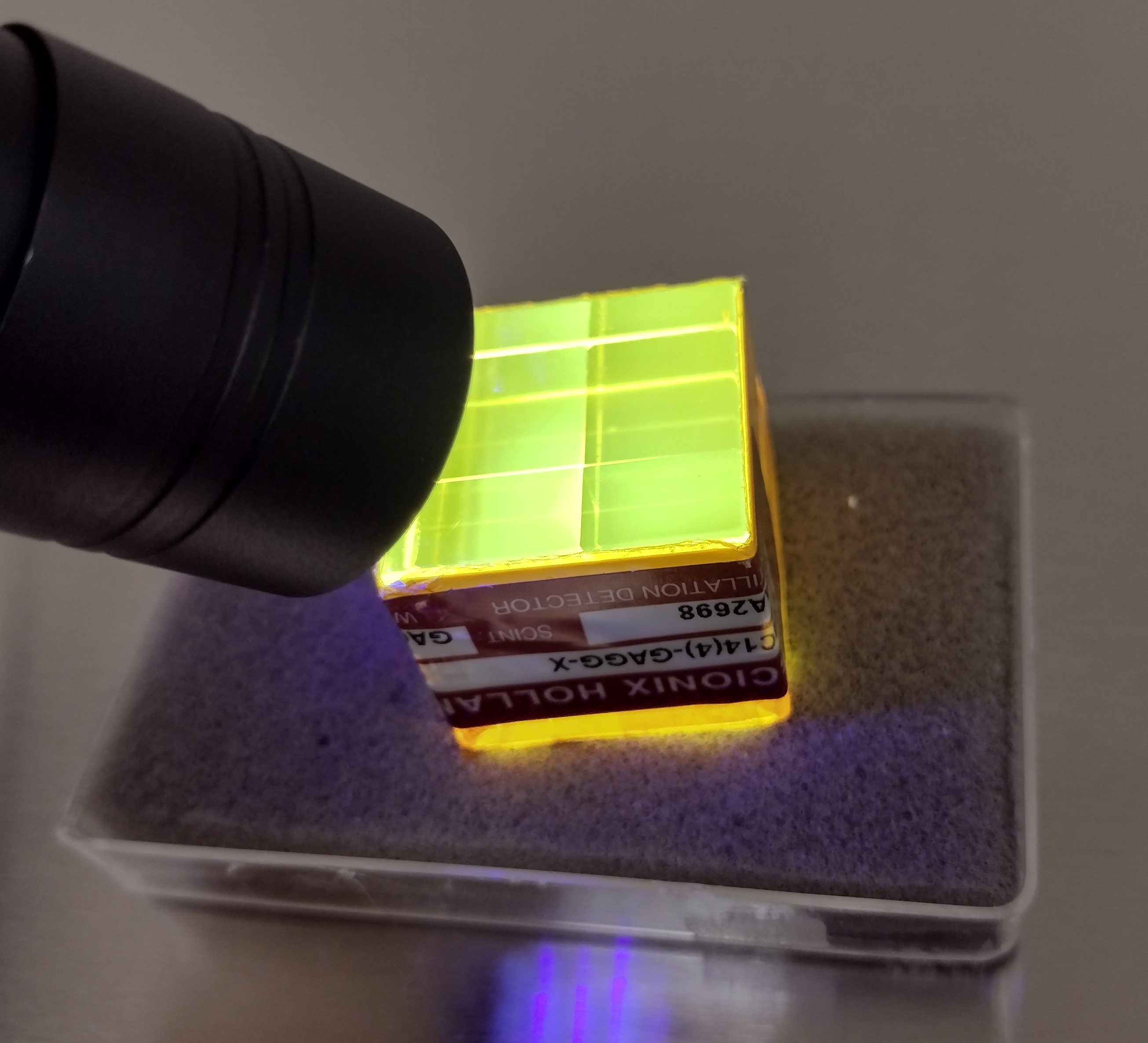}
    \captionof{figure}{The GAGG crystal used as the SNAPPY detector. Here it is illuminated by UV light which causes the crystal to produce scintillation light.}
    \label{FIG:GAGG}
\end{Figure}

\begin{Figure}
    \centering
    \includegraphics[width=0.85\linewidth]{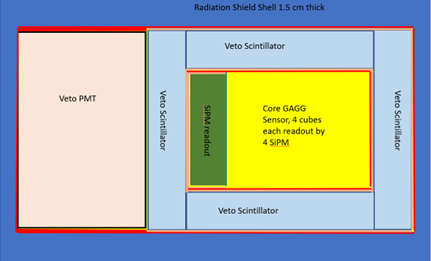}
    \captionof{figure}{Schematic of the detector (side-view). The detector is a nested design with the central GAGG detector being the active volume where the neutrino interaction occurs. The central detector is fully enclosed by reflective aluminum and has SiPM light sensors on one side. The outer detector is the veto plastic scintillator, which has a large PMT on one side and is used to reject charged particles that impinge on the central detector as they would also have to pass through the veto.}
    \label{FIG:DetectorSchematic}
\end{Figure}

Although the signature of cosmic-ray interactions in the GAGG detector is different from that of neutrino interactions, it is useful to suppress cosmic-ray interactions in the GAGG as much as feasible through both passive and active shielding. The outer passive shielding ideally consists of dense material of heavy nuclei, such as lead or steel. However, in a spacecraft for low earth orbit operation the eventual deorbiting and reentry into the atmosphere causes limitations. A monolithic metal block might not fully burn up on reentry, causing a risk of hitting the ground. Instead, SNAPPY uses a tungsten-loaded epoxy containing the largest fraction of tungsten powder that does not compromise the epoxy's structural integrity. Upon reentry the epoxy will easily burn up quickly and the remaining tungsten particles will be small enough to also burn up before hitting the ground.

The active shielding, the veto detector, is made of plastic scintillator read out by a photomultiplier tube (PMT). Charged particles will leave a pulse of light in this scintillator that can inhibit the trigger electronics, whereas neutrinos will not cause a veto-signal. Figure \ref{FIG:DetectorSchematic} shows the schematic of the detector assembly. Figure \ref{FIG:DetectorIntegration} shows the detector in front of the SNAPPY spacecraft.

\begin{Figure}
    \centering
    \includegraphics[width=0.85\linewidth]{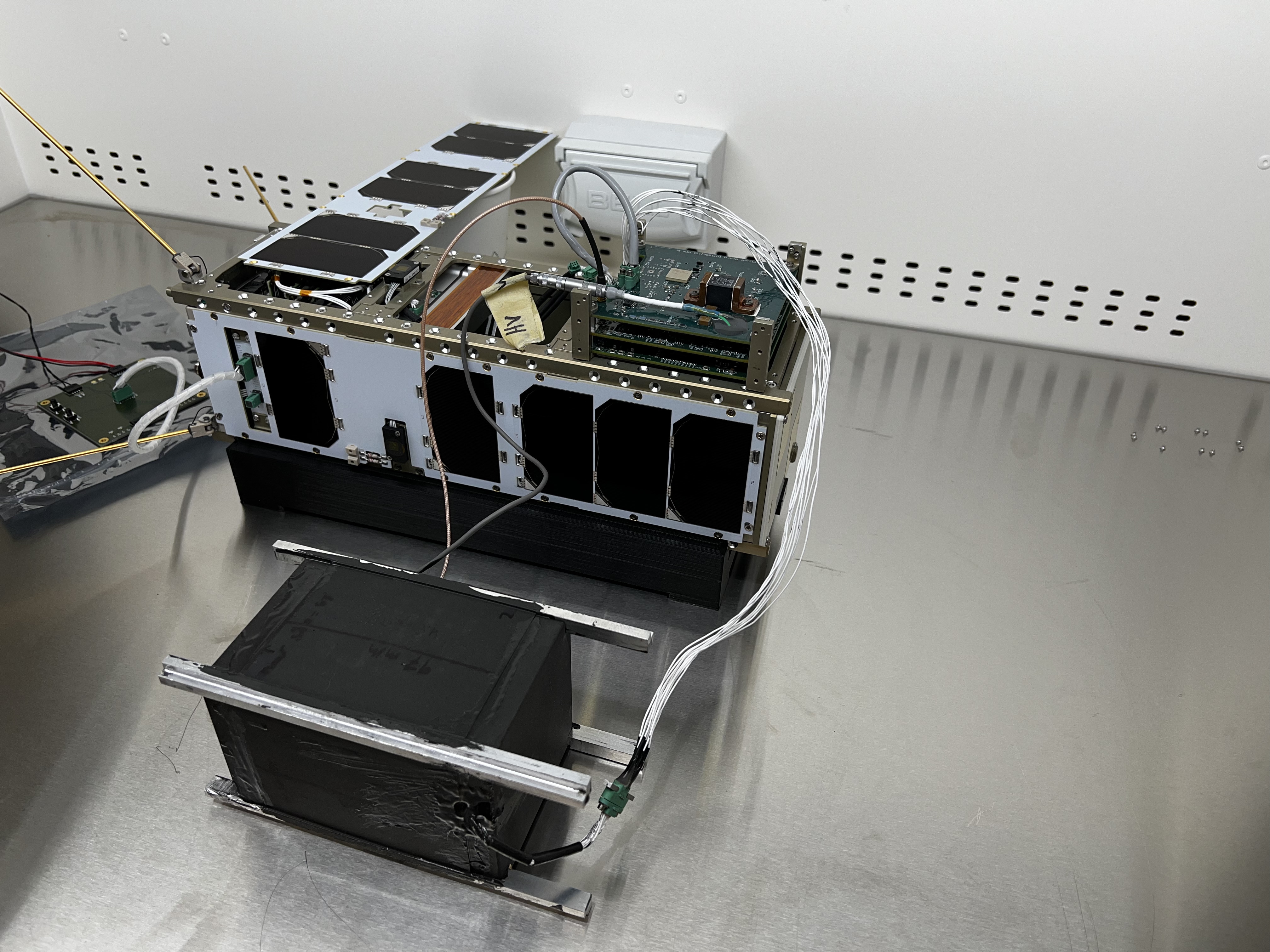}
    
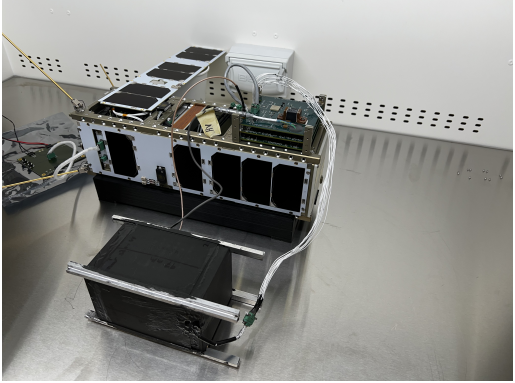
\captionof{figure}{The satellite during a typical ground operations moment before payload integration into CubeSat. The detector (or at least its outer shielding) is the black rectangular object in the foreground. The electronics stack designed by NASA MSFC is resting on the CubeSat frame.}
    \label{FIG:DetectorIntegration}
\end{Figure}

The total flux of cosmic rays outside the protective shield of the earth's magnetic field is known from previous measurements. Assuming cosmic rays are independent of one another, assival times are randomly (POisson) distributed in time. Two cosmic rays will rarely hit the SNAPPY detector (or even a larger future $\nu$SOL detector) close enough in time to mimic a neutrino interaction. However, the timing structure of cosmic rays has never been measured with sufficient timing resolution (in the range of tens of nano-seconds) to confirm this. For example, one might speculate that for some fraction of cosmic rays pairs of particles are the result of a common interaction that generates both, sending them in the same direction, but with slight delay, or else with momenta different enough such that over some travel distance a separation results. If this is the case, there might be situations when two cosmic rays traverse the detector in rapid succession. Even then they should not be confused for a neutrino as each should also leave light in the veto detector. But no detector has a detection efficiency of 100\%, and both cosmic rays might evade detection in the veto. Ultimately the rate of background events is undetermined until it is measured. That is the primary goal of SNAPPY.
\section{Mission Operations Planning}

The 3U CubeSat was launched in a Sun-Synchronous low-Earth (almost) circular orbit of approximately 500-km altitude. Such an orbit provides frequent but short opportunities for carrying out the necessary scientific experiments pertaining to our mission. This amounts to the detector being switched on when the spacecraft is over the North and the South poles of the Earth. At other points on the orbit, the detector is switched off and no data is collected. Such an orbit provides similar daily scientific data collection opportunities to being deployed in a geosynchronous transfer orbit, with the only difference being that the latter would have provided longer continuous (albeit infrequent) opportunities~\cite{NuSol_SFM25}. The CubeSat was launched utilizing a rideshare on the Exopod deployer of Exolaunch (see Figure~\ref{FIG:SNAPPY_Exopod}). Our main decision of using the SpaceX/Exolaunch services was driven by reasonable launch costs while also maximizing the number of launch opportunities for our CubeSat. In fact, among the various launch service providers who we contacted about potential ride share opportunities, Exolaunch provided us with the best cost per unit spacecraft mass alternative.

\begin{Figure}
    \centering
    \includegraphics[width=0.85\linewidth]{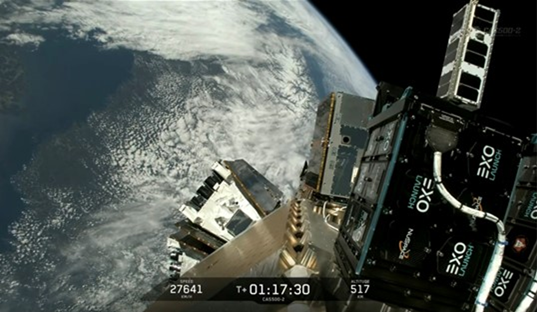}
    \captionof{figure}{SNAPPY being ejected from Exopod (photo obtained from SpaceX's livestream).}
    \label{FIG:SNAPPY_Exopod}
\end{Figure}

\par 
The CubeSat experiences significant eclipse times in the deployed orbit. The CubeSat has a battery to support limited operations during eclipses and to provide enhanced power (if needed) over the Sun-lit part of the orbit. The CubeSat has solar panels along four main sides as well as an extra deployable panel; the CubeSat attitude control system aims to maximize the power generated by the CubeSat by pointing the extra panel towards the Sun. Two of the main power consuming tasks for the CubeSat are communications with a ground station and data collection by the detector. If there is no power available for these tasks, the spacecraft defaults to a Sun-pointing mode. The spacecraft has two antennas onboard: UHF and S-BAND; however, owing to regulatory challenges, the UHF is not available for use during the mission. Hence, all communication tasks will be accomplished using the S-BAND antenna which allows for a maximum pointing error of $\pm 30^{\circ}$ for successful communication. Attitude control necessary for establishing communication links is accomplished by onboard 3-axis magneto-torquers. Because of limited control authority onboard (in contrast to having reaction wheels), it is not possible to maximize both Sun-pointing (for the extra panel) and ground station pointing (by the S-band antenna). However, we have a mechanism to determine a reference pointing direction that ensures communication link establishment without significant loss of power~\cite{Messick2}. This reference direction can be uploaded to the spacecraft through communication uplinks. Our previous studies had indicated that alternating between a Northern hemisphere ground station and a Southern hemisphere ground station seasonally (summer to winter and back) is the best strategy to satisfy communication and power constraints associated with our mission. To this end, our team has planned to use the services of Leaf Space and their ground stations in Iceland and Chile. At the time of launch, the desired pointing direction was set to be inertial, with the S-band antenna directly pointing towards the Iceland ground station when making a direct overhead pass.

\begin{Figure}
    \centering
    \includegraphics[width=0.85\linewidth]{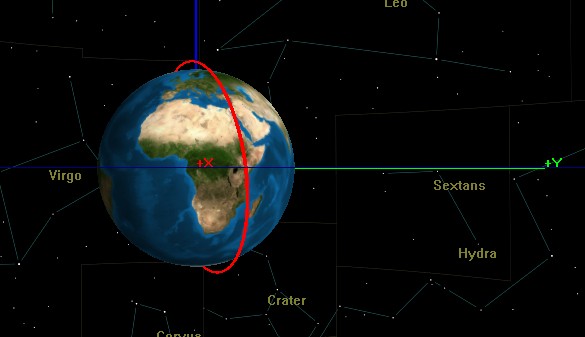}
    \captionof{figure}{SNAPPY orbit visualization based on post-deployment orbit parameters provided by SpaceX/Exolaunch.}
    \label{FIG:SNAPPY_Orbit}
\end{Figure}

Another unique challenge for our mission is the limited autonomy onboard. The CubeSat bus is the NanoAvionics built system with flight heritage. Ideally, one would want the flight software to autonomously switch the detector on/off based on its location determined by onboard GPS receiver. However, in order to keep the risks of the mission low, the only new component of the spacecraft is the detector itself. We therefore could neither make changes to the flight software, nor could make custom software add-on to the flight software for detector operation. Instead, we need to send in scripts for the payload controller during communication uplinks, to command the spacecraft to switch the detector on/off based at commanded times; these times are based on the predicted position of the CubeSat over a week-long window. This plan allows us to conform to existing communication protocol for CubeSats. These plans are crucial for in-space validation of the operation of the detector that is an enabling state for the larger $\nu$Sol mission~\cite{MESSICK2026420}.

\section{Data Parsing at Wichita State University}

To assist with cataloging data from SNAPPY, a database system was created on a local workstation at Wichita State University. Files are copied into a single folder known as the Launchpad, and a running daemon sorts and moves these files into various locations based off of their file extension. Such files involve data from the detector, scripts for the CubeSat, and telemetry data and logs from each subsystem. For some files such as detector files, they have to be parsed and turned from binary data into human-readable files to be used by CERN’s ROOT software, however others are simply just moved to their new locations. All files are then tracked by placing the information about their location and some metadata about them in a PostgreSQL database for easy accessibility to each dataset. Each file is given their own identifier, but files with the same name as an existing file are given the same identifier as the original with an incremented version number to group all duplicates together. 

With collaboration from WSU Aerospace Department, special timestamp files are created, sorted, and tracked that act as the backbone for mission scripts, and a special table exists to aggregate and catalog these mission plans into one location. It is with this information that a Python script will be used to generate the final scripts for the CubeSat and Mission Control.

\section{Orbit Analysis Tools}

On the computer that has been designated as the Mission Control Center, a Python script was created to pull TLE data from N2YO, a satellite database with information about each satellite including SNAPPY, and use the extracted TLE along with executing a helper script for NASA’s General Mission Analysis Tool (GMAT) version R2022a. The helper script shows the position of the CubeSat at the current time, generate one orbit in the past, three orbits in the future, and then use user-defined functions so that it will let GMAT track the current location of SNAPPY until the third orbit is complete; from then the original Python script will close GMAT and update the TLE, repeating the cycle until the script is forced to stop.
Potential ground stations are listed on the map as well as the location of the Mission Control Center. The MCS software is used to schedule communication with the spacecraft.

%Detector Functionality, Detector Results, Detector Simulations, Author: Brian
\section{Detector Functionality}

The Detector onboard SNAPPY is a nested detector consisting of a central detector composed of 4 GAGG (Gallium Aluminum Gadolinium Garnet) crystals, and an outer veto detector (Eljen plastic scintillator), both are good scintillators with sensitive light detectors attached. The central GAGG detector has a high density (6.7 g/cc) and a high component of gallium (22.57\% by mass), which is the target nucleus for solar neutrino interactions, and attached to one side of the 4 GAGG crystals is a quartz window that leads the scintillation photons into an array of 16 small SiPMs (Silicon PhotoMultipliers). The outer veto detector that fully surrounds the GAGG crystal is a light plastic scintillator with a large PMT (PhotoMultiplier Tube) on one side. This nested detector configuration allows the rejection of charged particles that penetrate the shielding, See Figure \ref{FIG:DetectorSchematic}.

The electronics were designed by engineers at NASA Marshall Space Flight Center and the associated University of Alabama in Huntsville. These engineers either worked directly on the ISS Advanced Neutron Spectrometer or closely with people who have, thus the delayed coincidence used for neutrino detection is very similar to the kind used for thermalized neutron detection and so the techniques, technology, and data structure are adapted from the ISS neutron detector. For the logic used by the FPGA to generate candidate neutrino events, see the following representation of the state machine (from NASA themselves).

\begin{Figure}
    \centering
    \includegraphics[width=0.85\linewidth]{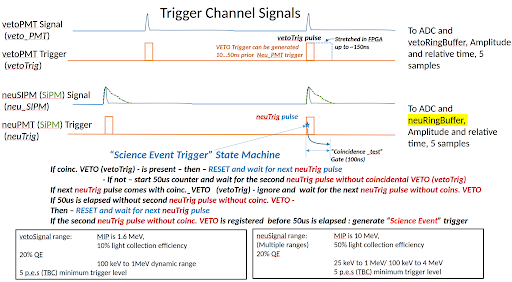}
    \captionof{figure}{Overview of the FPGA logic used to generate science events. Parameters are adjustable and can be changed in flight.}
    \label{FIG:FPGA_TriggerLogic}
\end{Figure}

The FPGA uses the above logic to generate “science events” when all of the conditions are satisfied. These science events contain a ring buffer of 5 ADC values (integrated signal over a gate with a 14-bit ADC) and 5 local timestamps representing 40 ns per unit of time; these pairs of ADC \& time values are for both the GAGG and veto. When a valid science event occurs, the ring buffer is dumped and will always have a time value of "1" in one of the 5 pairs of ADC \& time values for the GAGG; this was the event that triggered the saving of the ring buffer. These events are the neutrino candidate events and background events we are searching for. The other 4 ADC \& time pairs are not constrained by the actual science event trigger and these “nearby” events could be used for charged particle studies, namely space weather and galactic gamma ray burst studies (see detector simulation subsection).

\section{Ground Experiments}

The SNAPPY detector was calibrated and studied on the ground before launch using lab gamma ray sources such as Cs-137 and Co-60. The sources were placed on the exterior of the satellite roughly on the side with the detector, then the detector was put into a data taking mode. Then a linear calibration curve was made using the source data that converts measured ADC values into the energy deposited in the detector (for both veto and GAGG).
Below is a fitted histogram with data for a lab Na-22 see figure \ref{FIG:Na22_Gamma}. 

\begin{Figure}
    \centering
    \includegraphics[width=0.85\linewidth]{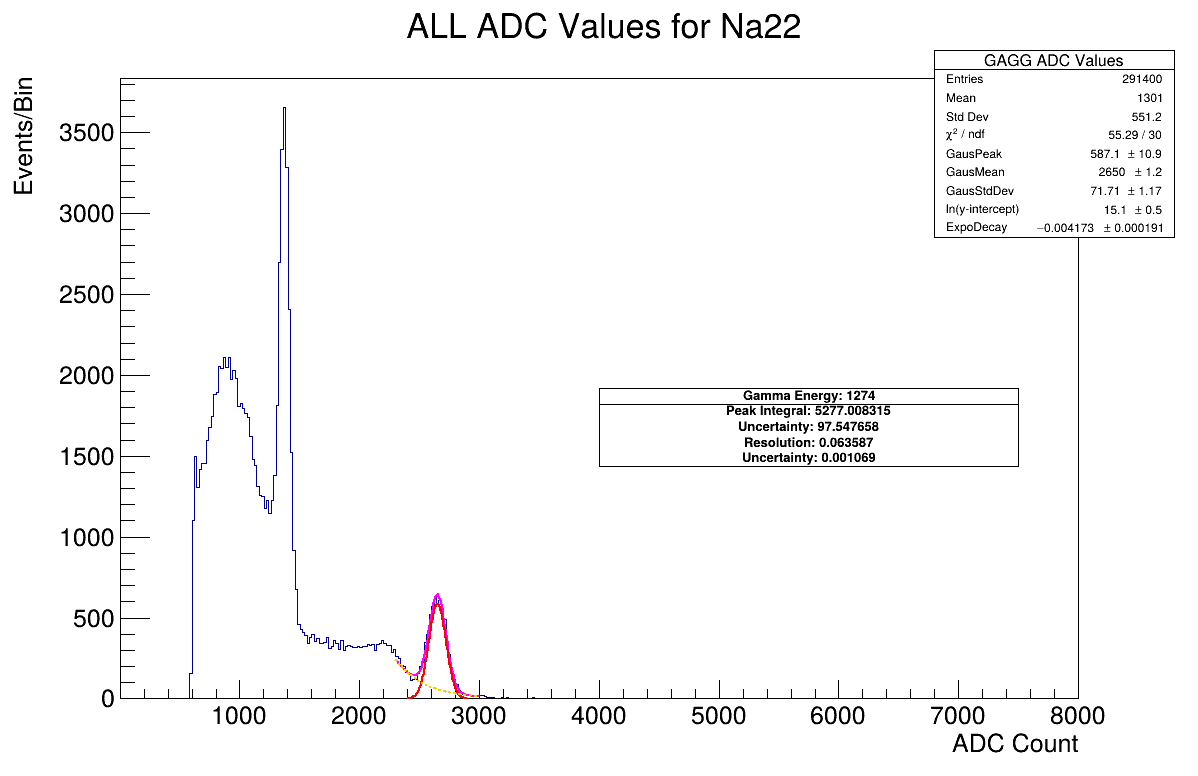}
    \captionof{figure}{SNAPPY GAGG detector results from Lab Na-22 source.}
    \label{FIG:Na22_Gamma}
\end{Figure}

Other runs were performed with the following sources: (Na-22, Cs-137, Mn-54, Zn-65, Ra-226, and Co-60) and the resulting calibration and resolution curves are shown in figures \ref{FIG:CalibrationCurveSimulation} \& \ref{FIG:ResolutionCurveSNAPPY}.

\begin{Figure}
    \centering
    \includegraphics[width=0.85\linewidth]{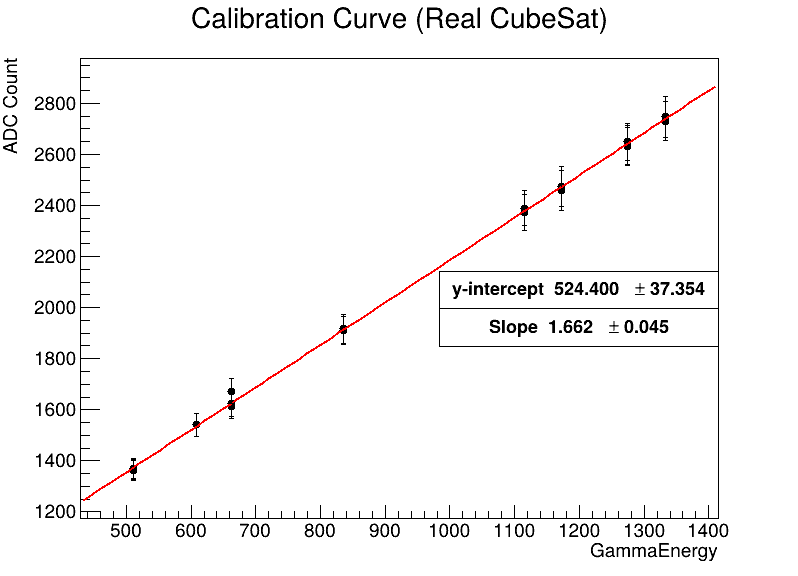}
    \captionof{figure}{Real SNAPPY data, GAGG ADC counts vs. Energy of incoming gamma rays vs. measured ADC values. The equation in the graph is the calibration used to convert between these values.}
    \label{FIG:CalibrationCurveSNAPPY}
\end{Figure}

\begin{Figure}
    \centering
    \includegraphics[width=0.85\linewidth]{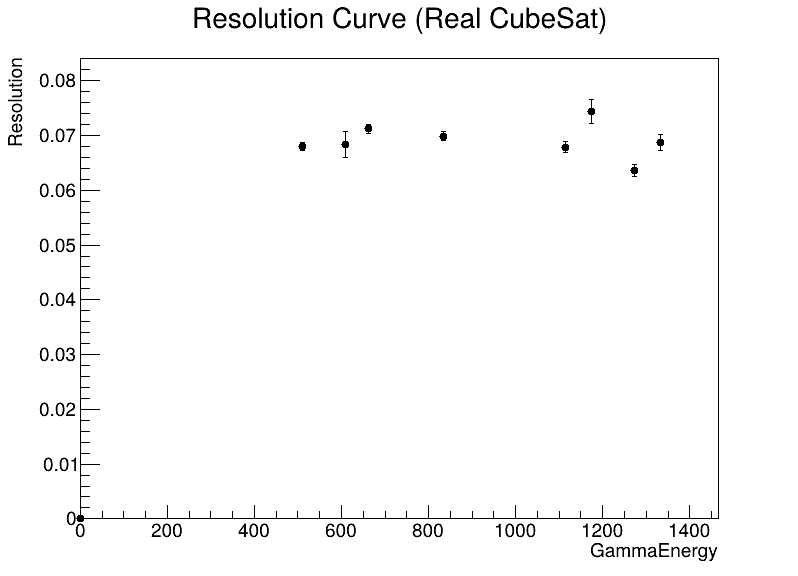}
    \captionof{figure}{The energy resolution of the GAGG detector. Notice that the energy resolution appears to be a constant 7\% due to the pulse shaping electronics.}
    \label{FIG:ResolutionCurveSNAPPY}
\end{Figure}

The veto detector is not designed for sharp energy resolution and the PMT is only on one side of the veto, which introduces geometric effects that are not present in the central detector. Regardless, a rough calibration study was performed with the veto so that a rough energy deposit can be estimated, which will feed into solar wind studies (see next subsection) that depend on an energy deposit estimate in both the GAGG and veto.
The real ground results can be used alongside Geant4 simulation results to most accurately characterize the delayed coincidence backgrounds in space. These calibration curves are also necessary for the space weather studies to reconstruct the energy spectrum of the incoming particles.

\section{Simulation Results for Gamma Ray Sources}

A custom Geant4 simulation was made of the SNAPPY detector and shielding for the analysis of the gamma ray detection efficiency of the detector and to find any other potential science that can be done with this unique detector and environment. The most promising secondary analysis performed in simulation was an analysis of the SNAPPY response to space weather.

\begin{Figure}
    \centering
    \includegraphics[width=0.85\linewidth]{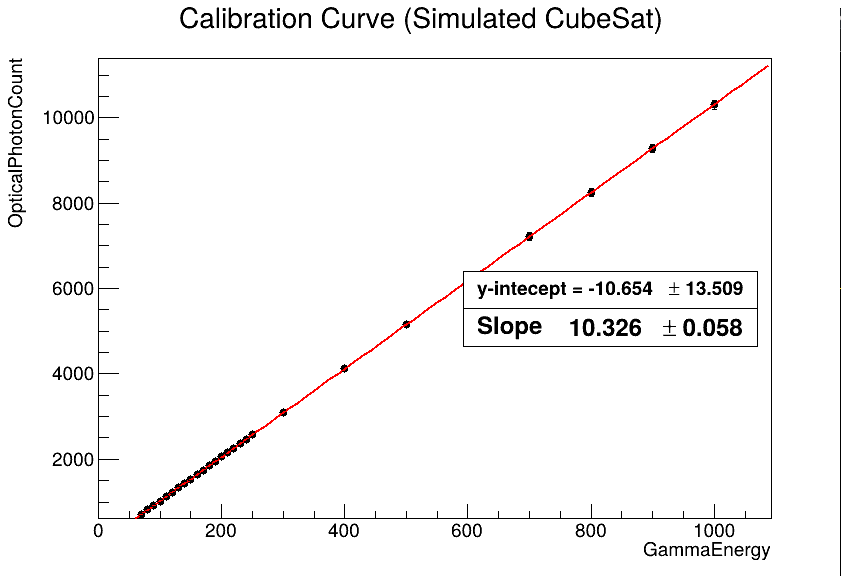}
    \captionof{figure}{Geant4 simulated optical photon count vs. gamma ray energy.}
    \label{FIG:CalibrationCurveSimulation}
\end{Figure}

\begin{Figure}
    \centering
    \includegraphics[width=0.85\linewidth]{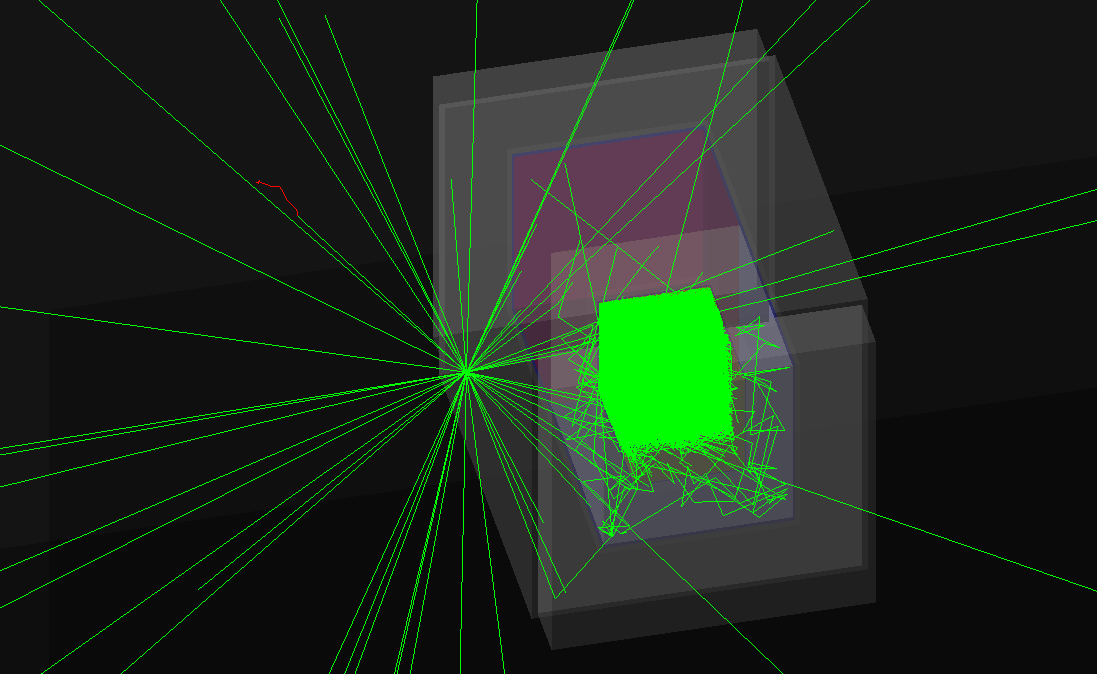}
    \captionof{figure}{Visualization of the Geant4 simulation that the project uses to predict and characterize the response of the detector to known sources and space weather. Here there is a low energy gamma ray source and a gamma happened to interact inside the central GAGG detector while bypassing the surrounding veto. The outermost grey volume is the tungsten-doped epoxy shielding, the innermost green volume is the GAGG detector with many light reflections occurring in the volume, and between these two regions is the veto volume with a large PMT attached on one side (in blue).}
    \label{FIG:GeantVis}
\end{Figure}

\section{Space Weather Simulations}

SNAPPY’s unique detector design allows for charged particle studies and therefore for space weather studies, so long as we look at the events in the ring buffer that aren’t the neutrino science event trigger. Preliminary studies were performed using Geant4 where an isotropic bath of cosmic rays were fired on the detector with a known energy spectrum. The photon counts measured by the simulation are directly proportional to the energy deposited in the detector and the ADC voltage readout by the real electronics boards. 

From these studies, SNAPPY can distinguish between the species of cosmic rays and Solar Energetic Particles (SEP) such as protons and alpha particles by using the GAGG+veto ADC counts. By comparing the energy deposited in both detectors, and tracing which particle deposited that energy by using simulation generator information, we can build a map of which regions in the GAGG+veto ADC count plane correspond to which particles. See the figure below.

\begin{Figure}
    \centering
    \includegraphics[width=0.85\linewidth]{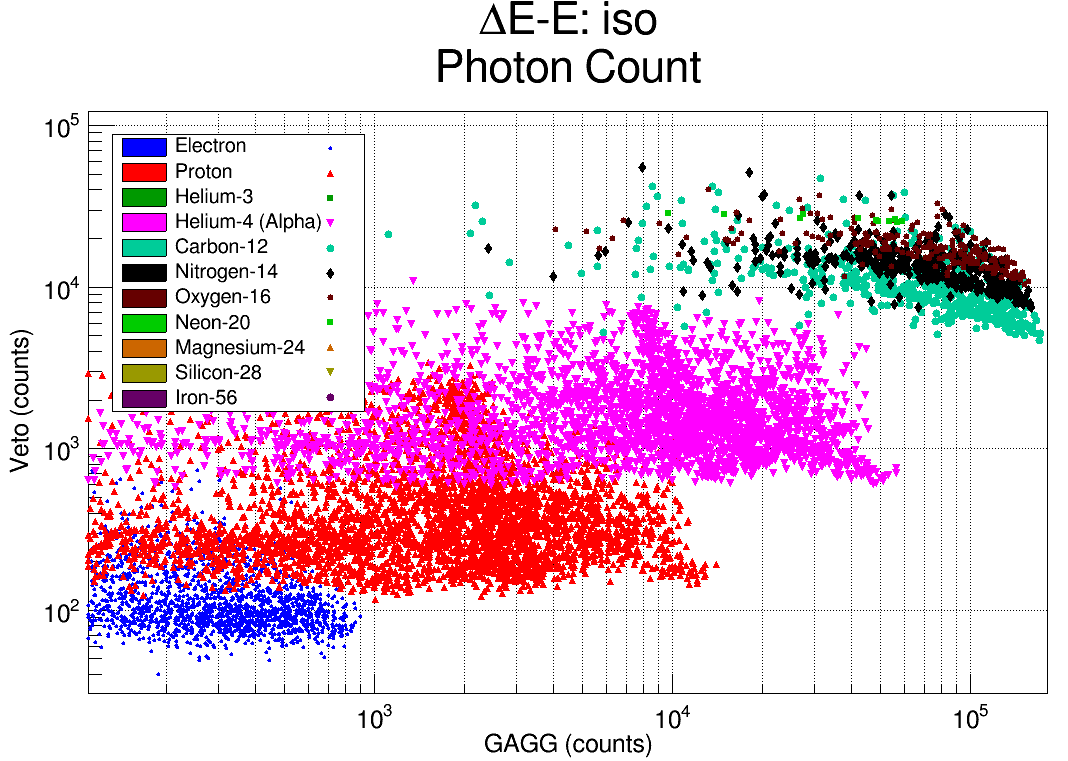}
    \captionof{figure}{Geant4 simulation data with number of optical photons counted for each detector on each axis. This data shows how most of the measured photon count space has distinct regions corresponding to certain particles that are unmistakable (confidence level >95\% of being one particular species of cosmic ray or SEP) and the remainder of the space can be broken into regions where we can estimate the portion of particles that belong to each species.}
    \label{FIG:EdE}
\end{Figure}

\section{Conclusion}

SNAPPY is a technology demonstration mission of the $\nu$SOl technology, resolving solar neutrino signals over the backgrounds by going over the Van Allen belts at the north and south poles and detecting events using the same Ga-71 delayed coincidence measurement technique. The custom data processing pipeline will automatically decode and archive the data. The team will analyze the data to identify space weather and to characterize the false-positive neutrino candidate events. The $\nu$SOL collaboration will use the results to plan the design of a solar probe neutrino detector mission to the sun.

\bibliography{bib.bib}
\bibliographystyle{unsrt}

%%%%%%%%%%%%%%%%%%
% \clearpage
% \input{Instructions} %comment this line out before downloading PDF for submission
%%%%%%%%%%%%%%%%%%

\end{multicols*}

\end{document}